\documentclass[aps,prl,superscriptaddress,unsortedaddress,twocolumn,preprintnumbers,amsmath,amssymb]{revtex4-1}
\usepackage{graphicx}% Include figure files
\usepackage{dcolumn}% Align table columns on decimal point
\usepackage{bm}% bold math
\usepackage{upgreek}
\usepackage{txfonts}
\usepackage{color}
\newsavebox{\tempbox}

\begin{document}

\title{Josephson junctions in a local inhomogeneous magnetic field}

%\author{R. A. Hovhannisyan$^{1,2}$}
%\author{O. M. Kapran$^1$}
%\author{T. Golod$^1$}
%\author{S. Yu. Grebenchuk$^{1,2}$}
%\author{V. V. Dremov$^2$}
%\author{V. S. Stolyarov$^2$}
\author{V. M. Krasnov$^{1,2}$}
\email{Vladimir.Krasnov@fysik.su.se}

\affiliation{$^1$ Department of Physics, Stockholm University,
AlbaNova University Center, SE-10691 Stockholm, Sweden }
\affiliation{$^2$ Moscow Institute of Physics and Technology,
State University, 9 Institutsiy per., Dolgoprudny, Moscow Region
141700 Russia}

\date{\today}

\begin{abstract}
A Josephson junction can be subjected to a local, strongly
inhomogeneous magnetic field in various experimental situations.
Here this problem is analyzed analytically and numerically. A
modified sine-Gordon type equation in the presence of
time-dependent local field  is derived and solved numerically in
static and dynamic cases. Two specific examples of local fields
are considered: induced either by an Abrikosov vortex, or by a tip
of a magnetic force microscope (MFM). It is demonstrated that
time-dependent local field can induce a dynamic flux-flow state in
the junction with shuttling, or unidirectional ratchet-like
Josephson vortex motion. This provides a mechanism of detection
and manipulation of Josephson vortices by an oscillating MFM tip.
In a static case local field leads to a distortion of the critical
current versus magnetic field, $I_c(H)$, modulation pattern. The
distortion is sensitive to both the shape and the amplitude of the
local field. Therefore, the $I_c(H)$ pattern carries information
about the local field distribution within the junction.
%We argue that such a distortion can be employed for reconstruction of the spatial distribution of the field.
This opens a possibility for employing a single Josephson junction
as a scanning probe sensor with spatial resolution not limited by
its geometrical size, thus obviating a known problem of a
trade-off between field sensitivity and spatial resolution of a
sensor.

\end{abstract}

\pacs{
%%74.72.Hs, %Bi-based cuprates
%%74.78.Fk, %Multilayers, superlattices, heterostructures
74.50.+r, %Tunneling phenomena; point contacts, weak links, Josephson effects
85.25.Cp %Josephson devices
}

\maketitle

\section{I. Introduction}

Properties of Josephson junctions (JJ's) with spatially uniform
parameters in a homogenous magnetic field are well studied
\cite{Barone}. Also, spatially nonuniform JJ's in a homogeneous
field were considered earlier
\cite{Vasenko_1981,Geshkenbein_1992,Scott_1996,Krasnov_1997,Costabile_2001}.
However, in many experimental situations JJ's are subjected to a
local, strongly inhomogeneous magnetic field. For example, it can
originate from a self-induced flux in JJ's with a sign-reversal
order parameter
\cite{VanHarlingen_1995,Tsuei_2000,Hilgenkamp_2003}; appear in
JJ's containing ferromagnetic interlayers with spatially
inhomogeneous thicknesses
\cite{Koshelev_2003,Goldobin_2008,Goldobin_2012,Bakurskiy_2016},
nanoparticles or domain walls
\cite{Weides_2008,Birge_2012,Iovan_2014,Golovchanskiy_2016,Iovan_2017,Aarts_2017};
in JJ's with a local current injection
\cite{Gaber_2005,Milosevic_2009}; can be induced by a nearby
Abrikosov vortex
\cite{Finnemore_1994,Golod_2010,Golod_2015,Golod_2019b}, by a
sharp tip of a Magnetic Force Microscope (MFM)
\cite{Budakian_2019,Dremov_2019}, etc. Although such a situation
has been considered previously, often this has been done without
proper substantiation. To my knowledge, there is no established
formalism for a general treatment of such a problem, especially in
the dynamic case.

Another motivations of this work is related to a recent proposal
to use a single planar JJ as a scanning probe sensor
\cite{Golod_2019a}. The leading superconducting scanning probe
technique today is the scanning SQUID (superconducting quantum
interference device) microscopy \cite{Kirtley_2016,Zeldov_2016}.
Despite many advantages, SQUID's suffer from a trade-off problem
between field sensitivity and spatial resolution. SQUID's, as well
as most other superconducting magnetic sensors, are measuring flux
with a resolution $\delta\Phi$ determined by the flux quantum
$\Phi_0$. Therefore, field sensitivity is inversely proportional
to the sensor (pickup loop) area $S$, $\delta H =\delta \Phi/S$.
On the other hand, spatial resolution is determined by the sensor
size $\delta x \sim S^{1/2}$. Consequently, the better is spatial
resolution, the worse is field sensitivity, $\delta H \sim
1/\delta x^2$. In Ref. \cite{Golod_2019a} it was argued that a
sensor based on a single planar JJ would be able to obviate the
trade-off problem at least in one spatial direction. Similar to a
SQUID, the field sensitivity of a planar JJ is also inversely
proportional to the area. Therefore, obviation of the trade-off
problem would require independence of spatial resolution on the
junction size. That is, the junction should be able to resolve
spatial variation of magnetic field at a scale significantly
smaller than the junction length. This brings us again to the
problem of a JJ in a local spatially inhomogeneous magnetic field.

In this work I consider analytically and numerically a response of
a single JJ to a local inhomogeneous and time-dependent magnetic
field. First, a modified sine-Gordon equation for this case is
derived. The equation is then solved numerically both for short
and long junctions, and both in static and dynamic cases. Two
specific examples (without loosing generality) are considered with
a local field induced either by an Abrikosov vortex, or by a tip
of MFM. It is demonstrated that a time-dependent local field can
induce a dynamic flux-flow phenomenon with either shuttling or
unidirectional ratchet-like motion of Josephson vortices in the
junction, which provides a mechanism for detection of Josephson
vortices by MFM \cite{Dremov_2019}. Analysis of the static case
shows how the critical current versus magnetic field, $I_c(H)$,
modulation patterns are distorted with introduction of the local
inhomogeneous field. Importantly, the shape of distorted $I_c(H)$
patterns depends both on the shape, amplitude and position of the
local field. Therefore, the $I_c(H)$ pattern carries detailed
information about the local field and it should be possible to
extract field distribution within the junction using proper
mathematical treatment. This would open a possibility for making a
scanning probe sensor based on a single planar Josephson junction
with spatial resolution not limited by its geometrical size, thus
obviating the trade-off problem between sensitivity and
resolution.

\section{II. Theoretical analysis}

\begin{figure}[t]
    \includegraphics[width=0.4\textwidth]{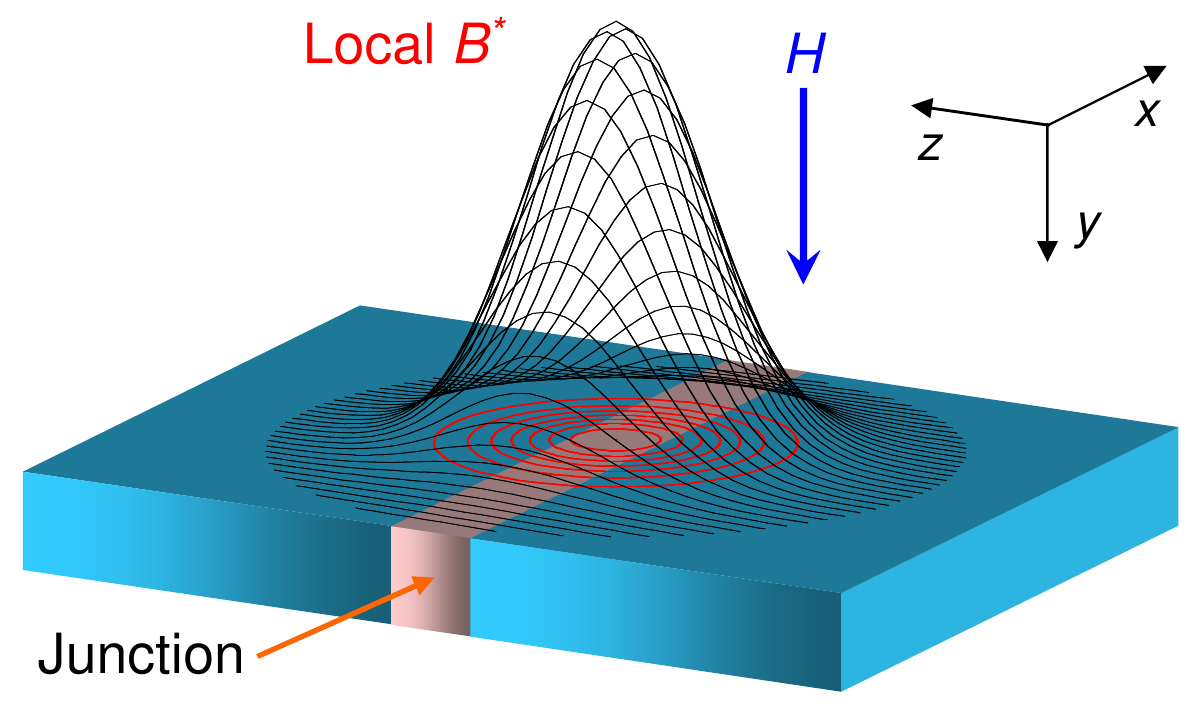}
     \caption{\label{fig:1} (Color online). Sketch of a Josephson junction in a local inhomogeneous magnetic field $B^*$.}
\end{figure}

Figure \ref{fig:1} represents a sketch of the studied problem. Let
us consider a Josephson junction in applied uniform magnetic field
$H$ in $y$-direction and a local nonuniform field $B^*$. The bias
current is applied from one electrode to another in $z$-direction.
I start with derivation of transport equations. Although main
parts of the derivation are well known, I will show it in some
details both for the sake of pedagogical completeness and because
sometimes different equations have been introduced in literature
without proper substantiation.

According to the two-fluid model, current through a Josephson
junction has superconducting and quasiparticle (normal)
components. The supercurrent density is given by the DC-Josephson
relation,
\begin{equation}
J_s = J_{c0} \sin (\varphi) \label{Is},
\end{equation}
where $J_{c0}$ is the critical current density at zero magnetic
field and $\varphi$ is the Josephson phase difference. The
quasiparticle current density is equal to $J_n = V/R_n A$, where
$V$ is voltage across the junction, $R_n$ is the normal resistance
of the junction and $A$ is the junction area. Using the
AC-Josephson relation $V=(\Phi_0/2\pi c) (\partial \varphi
/\partial t)$, where $c$ is the speed of light in vacuum, it can
be written as
\begin{equation}
J_n = \frac{\Phi_0}{2\pi c R_n A}\frac{\partial\varphi}{\partial t} \label{In}.
\end{equation}
Note that for planar JJ's the junction area $A$ (in the $(x,y)$-plane, see Fig. \ref{fig:1}) is much smaller than the sensor area $S$ (in the $(x,z)$-plane) \cite{Golod_2019a,Boris_2013}.
Using Maxwell equation $rot{B} = \frac{4\pi}{c}J
+\frac{1}{c}\frac{\partial D}{\partial t} $ we can write the total current through the
junction %(in $z$-direction for the considered geometry, see Fig. \ref{fig:1})
as:
\begin{equation}
J_z = \frac{c}{4\pi} (\frac{\partial B_y}{\partial
x}-\frac{\partial B_x}{\partial y})-\frac{1}{4\pi} \frac{\partial D}{\partial t}. \label{Maxwell}
\end{equation}
Here the last term represents the displacement current density
\begin{equation}
J_d = \frac{\Phi_0 C}{2\pi c A}\frac{\partial^2\varphi}{\partial
t^2}, \label{Id}
\end{equation}
where $C$ are the junction capacitance.

Both $x$ and $y$ components of magnetic induction in Eq.
(\ref{Maxwell}) are essential. The $x$-component, parallel to the
junction, has contributions from the bias current, $B_{xb}$, and
the local nonuniform field $B^*_x$, $B_x=B_{xb}+B^*_x$. The bias
contribution is related to the bias current density in the
$z$-direction, $J_b$, as
\begin{equation}\label{Ibias}
\frac{\partial B_{xb}}{\partial y} = - \frac{4\pi}{c} J_{b}.
\end{equation}
This is how the bias term, which plays the role of the driving
force for junction dynamics, enters the sine-Gordon equation.

From Eqs. (\ref{Is}-\ref{Ibias}) we obtain
\begin{equation}
J_d+J_n+J_s = \frac{c}{4\pi} (\frac{\partial B_y}{\partial
x}-\frac{\partial B^*_x}{\partial y})+J_b. \label{Itot}
\end{equation}

The $y$-component of magnetic field, going through the junction, induces a phase gradient in the junction,
\begin{equation}
\frac{\partial \varphi}{\partial x}=\frac{2\pi d_{eff}}{\Phi_0}
B_y. \label{dfdx}
\end{equation}
Here $d_{eff}$ is the so-called magnetic thickness of the
junction. $B_y$ is the total (screened) induction in the junction
subjected both to the applied uniform field $H$ and the local
nonuniform field $B^*$. It is generally not known and should be
determined. To do so we separate the phase shift $\varphi^*$
caused solely by $B^*_y$.
\begin{eqnarray}
\varphi=\phi+\varphi^*,\\
\frac{\partial \varphi^*}{\partial x}=\frac{2\pi d_{eff}}{\Phi_0}
B^*_y,\\
\frac{\partial \phi}{\partial x}=\frac{2\pi d_{eff}}{\Phi_0}
(B_y-B^*_y). \label{df12}
\end{eqnarray}
$\varphi^*(x)$ is a known function, determined (up to an integration constant) by integration of Eq. (9) along the junction length.

Using Eqs. (\ref{dfdx}-\ref{df12}) we can write
\begin{equation}
J_d+J_n+J_s = \frac{c\Phi_0}{8\pi^2 d_{eff}} \frac{\partial^2
\phi}{\partial x^2}+\frac{c}{4\pi}(\frac{\partial B^*_y}{\partial
x}-\frac{\partial B^*_x}{\partial y}) + J_b. \label{Im3}
\end{equation}
Note that the second term in the right-hand-side represents $z$-component of $rot B^*$. Since $B^*$ does not induce vacuum
currents, $rot B^*=0$, this term vanishes. Substituting Eqs.
(1,2,4) in Eq. (11) we obtain the desired modified sine-Gordon-type equation:
%\begin{equation}
%\frac{c\Phi_0}{8\pi^2 d_{eff}} \frac{\partial^2 \phi}{\partial
%x^2} - \frac{\Phi_0 C}{2\pi A}\frac{\partial^2\varphi}{\partial
%t^2}- \frac{\Phi_0}{2\pi R_n A}\frac{\partial\varphi}{\partial t}
%= J_{c0} \sin (\phi+\varphi^*)-J_b, \label{SGmod}
%\end{equation}

\begin{equation}
\frac{\partial^2 \phi}{\partial \tilde{x}^2} -
\frac{\partial^2\phi}{\partial \tilde{t}^2}-
\alpha\frac{\partial\phi}{\partial \tilde{t}}  = \sin
(\phi+\varphi^*)-j_b + \frac{\partial^2\varphi^*}{\partial
\tilde{t}^2}+ \alpha\frac{\partial\varphi^*}{\partial \tilde{t}}.
\label{SGnorm}
\end{equation}
Here space, $\tilde{x}=x/\lambda_J$, is normalized by the
Josephson penetration depth,
$\lambda_J=\sqrt{\frac{c\Phi_0}{8\pi^2 d_{eff} J_{c0}}}$ and time,
$\tilde{t}=\omega_p t$, by the inverse plasma frequency
$\omega_p^{-1} =  \sqrt{\frac{\Phi_0 C}{2\pi c A J_{c0}}}$,
$\alpha= (\omega_p R_n C)^{-1}$ is the damping parameter and
$j_b=J_b/J_{c0}$.
%For slowly changing local field $B^*$ we can neglect the two time-dependent  $\varphi^*$ terms in the r.h.s. of Eq. (\ref{SGnorm}) and use a simplified equation:
%\begin{equation}
%\frac{\partial^2 \phi}{\partial \tilde{x}^2} - \frac{\partial^2\phi}{\partial \tilde{t}^2}- \alpha\frac{\partial\phi}{\partial \tilde{t}}  = \sin (\phi+\varphi^*)-j_b. \label{SGsimp}
%\end{equation}
%The time evolution of $B^*$ should be slow with respect to the Josephson frequency.
%%, which is usually the case.
%For example in the MFM experiment the tip frequency is about 80 kHz, while the Josephson frequency is in the range
Eq. (\ref{SGnorm}) should be solved with respect to $\phi$ for the
known $\varphi^*(x,t)$ with boundary conditions at the junction
edges $x=0, ~L_x$:
\begin{equation}
\frac{\partial \phi}{\partial \tilde{x}}(0,L_x)=\frac{2\pi
d_{eff}\lambda_J}{\Phi_0} H. \label{BC}
\end{equation}
Note that thanks to separation of variables, Eq. (8), the local
nonuniform field drops out from the boundary conditions. This
occurs because at the junction edges $B_y(0,L_x)=H+B^*_y(0,L_x)$
so that $B_y(0,L_x)-B^*_y(0,L_x)=H$, which together with Eq. (10)
yields Eq. (\ref{BC}).

In what follows we will normalize magnetic field by
$H_0=\Phi_0/\pi \Lambda_1\lambda_J$ and voltage by
$V_0=\Phi_0\omega_p/2\pi c$, where
$\Lambda_1=t_i+\lambda_1\coth(d_1/\lambda_1)+\lambda_2\coth(d_2/\lambda_2)$,
$t_i$ is the junction interlayer width, $d_{1,2}$ are the widths
and $\lambda_{1,2}$ are the London penetration depths of the two
electrodes (for details see Ref. \cite{Krasnov_2001}).

\section{III. Static case}

In the static case the only current component is the supercurrent
$J_s$, Eq. (\ref{Is}), and Eq. (\ref{SGnorm}) is reduced to
\begin{equation}
\frac{\partial^2 \phi}{\partial \tilde{x}^2} = \sin
(\phi+\varphi^*)-j_b. \label{SGstat}
\end{equation}
According to Eqs. (8-10), the local supercurrent density $J_s(x)$
directly depends on the local field $B^*(x)$. Experimentally
measurable quantity, however, is the critical current $I_c$, which
represents the maximum value of the integral of $J_s$ along the
junction length.
%\begin{equation}\label{Ic}
%I_c=\int^{L_x}_0 { J_s dx}.
%\end{equation}
The value of $I_c$ alone does not disclose the $B^*(x)$
distribution. However, as we will show below, the $I_c(H)$
modulation pattern does carry information about distribution
of magnetic induction in the junction.

\begin{figure}[t]
    \includegraphics[width=0.3\textwidth]{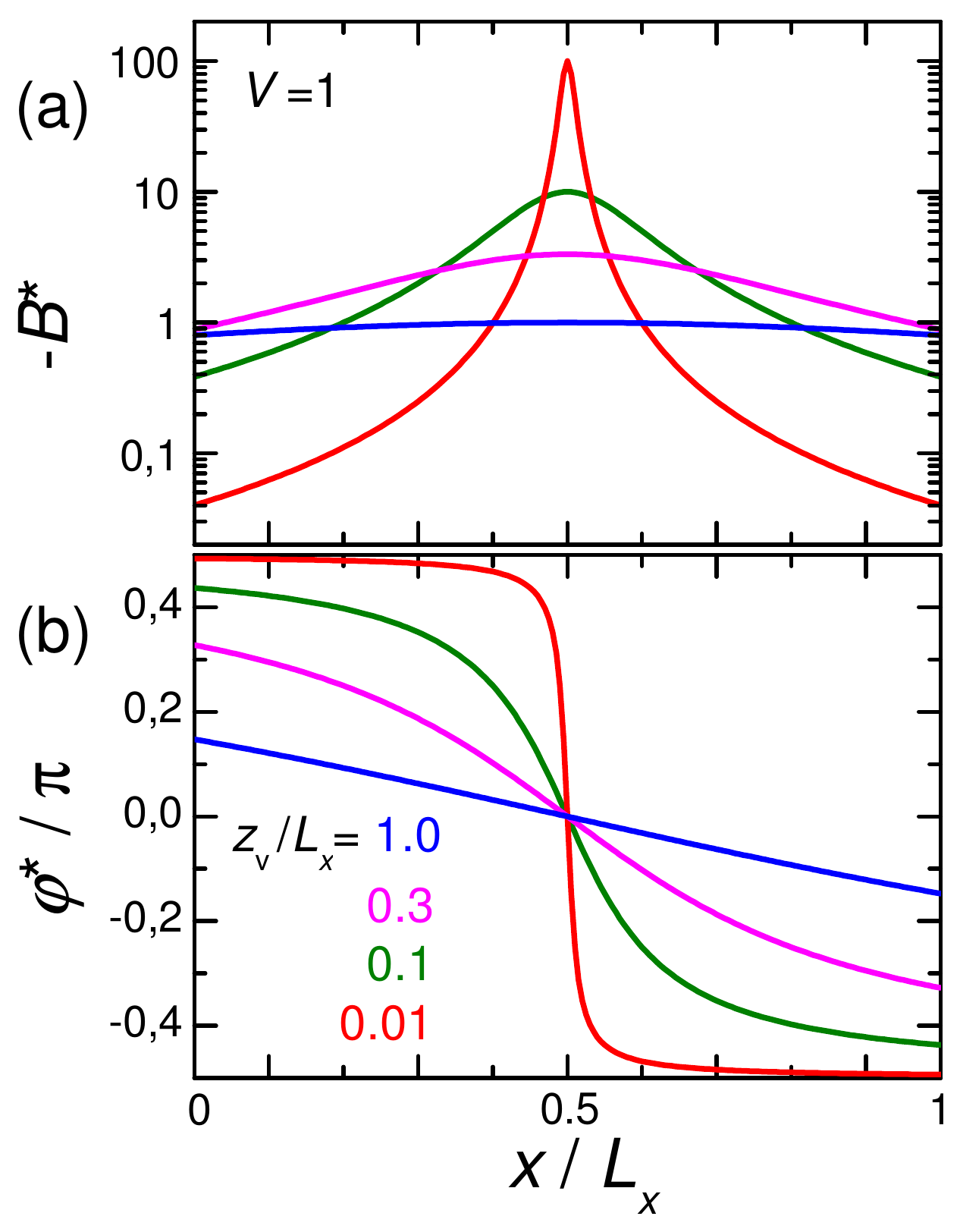}
     \caption{\label{fig:2} (Color online). (a) Abrikosov vortex-induced stray
     fields, $-B^*(x)$, at four different distances $z_v$ of the vortex to the junction and at $x_v=0.5 L_x$  (in the semi-logarithmic
     scale). The minus sign is due to the opposite sign of the vortex stray field with respect to the
     vortex. (b) Corresponding Josephson phase shifts $\varphi^*(x)$.}
\end{figure}

\subsection{III A. Short junctions}

First we consider the simplest case of a short junction
$L_x\lesssim \lambda_J$. In this case we may neglect magnetic
field screening in the junction, i.e. set the second derivative
term in the left-hand side of Eq. (14) to zero. From Eq.
(\ref{BC}), we explicitly obtain $\phi(x)\simeq (2\pi d_{eff}
H/\Phi_0) x + \phi_0$, where $\phi_0$ is the integration constant.
The total supercurrent is calculated directly by integration of
$\sin[\phi(x)+\varphi^*(x)] dx$. The critical current is obtained
by maximization with respect to the
integration constant $\phi_0$.

To demonstrate how local inhomogeneous field distorts the $I_c(H)$
pattern we consider the case when the local field is created by
stray fields from an Abrikosov vortex. This case has been
described in details in a recent work \cite{Golod_2019b}, in which
it was shown that vortex-induced Josephson phase shift is well
described by the equation:
\begin{equation}\label{AQ1}
\varphi^*(x)= -V \arctan \left(\frac{x-x_v}{|z_v|}\right),
\end{equation}
where $V$ is the vorticity (+1 for a vortex, -1 for an
antivortex), $x_v$ is the coordinate of the vortex along the
junction and $z_v$ is the distance to the junction. Figure
\ref{fig:2} shows (a) vortex stray fields, $-B^*(x)$, in the
junction and (b) corresponding Josephson phase shifts
$\varphi^*(x)$ for an Abrikosov vortex, $V=1$, at four different
distances $z_v$ to the JJ along the junction middle line $x_v=0.5
L_x$. The closer is the vortex to the junction, the sharper and
larger is the local stray field $B^*(x)$. Note that the sign of
the stray field is opposite to that in the vortex, leading to the
minus sign in Eq. (15) \cite{Golod_2019b}.

\begin{figure*}[t]
    \includegraphics[width=0.98\textwidth]{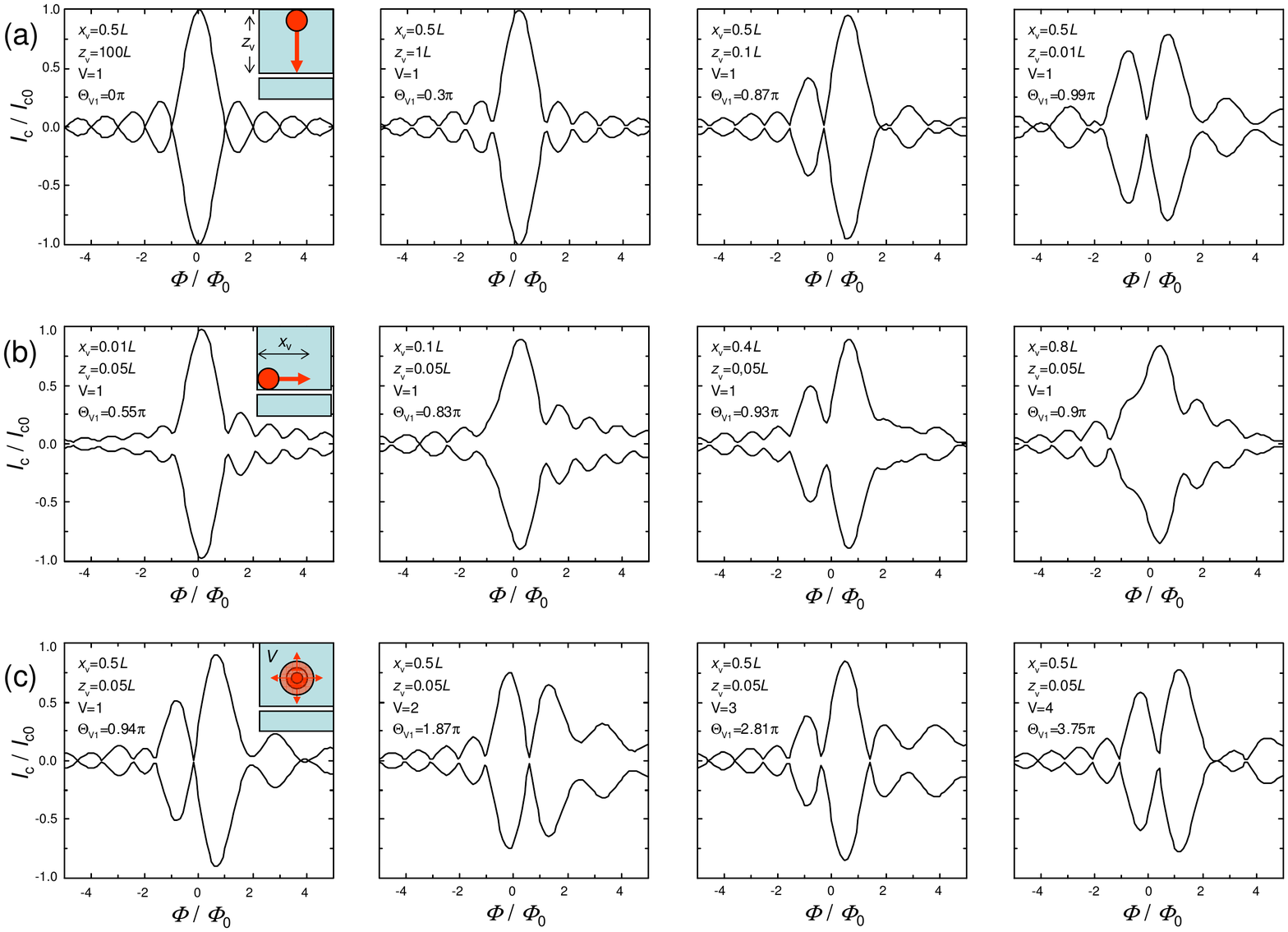}
     \caption{\label{fig:3} (Color online). Calculated  $I_c(H)$ dependencies for a short junction with a static local field induced by stray fields from an
     Abrikosov vortex at different (a) distances, (b) positions and (c) vorticities, as indicated in the insets.
     Panels in row (a) from left to right demonstrate a progressive distortion of $I_c(H)$ patterns upon approaching of the vortex to the junction along the middle line $x_v=0.5 L_x$.
     Panels in row (b) - upon moving of the vortex from the left to the right edge of the junction at constant distance $z_v=0.05 L_x$. Panels (c) show evolution of $I_c(H)$ patterns
     with changing the vorticity $V$ at a fixed vortex position $x_v=0.5 L_v$, $z_v=0.05 L_x$. Note that each $I_c(H)$ pattern is clearly distinguishable.
     This demonstrates that the effect of local inhomogeneous magnetic field is uniquely encoded in the shape of the $I_c(H)$ pattern. Therefore, it should be possible
     to reconstruct spatial distribution $B^*(x)$ from the analysis of $I_c(H)$ modulation. }
\end{figure*}

Figure \ref{fig:3} illustrates an evolution of $I_c(H)$ patterns
with changing local fields: row (a) upon approaching the vortex to
the junction along the middle line $x_v=0.5 L_x$; row (b) upon
moving the vortex from the left to the right sides of the junction
at $z_v=0.05 L_x$; and row (c) upon increasing the vorticity for a
fixed position $x_v=0.5 L_x$, $z_v=0.05 L_x$. The quantity
$\Theta_{v1}$ represents the absolute value of the total
vortex-induced Josephson phase shift
$\Theta_{v1}=|\varphi^*(L_x)-\varphi^*(0)|$, which yields the
total local flux in the junction,
$|\Phi^*|=(\Theta_{v1}/2\pi)\Phi_0$.

The left panel in row (a) represents the case with far away
vortex, $z_v=100 L_x$, and negligible local field, $B^*\simeq 0$.
In this case $I_c(H)$ follows the standard Fraunhofer pattern.
Figs. \ref{fig:3} (a) demonstrate that the $I_c(H)$ patterns get
progressively distorted upon approaching the vortex to the JJ,
accompanied by a subsequent sharpening and increasing of $B^*$, as
shown in Fig. \ref{fig:2} (a). Figs. \ref{fig:3} (b) correspond to
the case when only the position of $B^*$ maximum, $x_v$, is
changes, while the amplitude and the shape of $B^*$ remains the
same. In Figs. \ref{fig:3} (c) the position and the shape remain
the same and only the amplitude of $B^*$ changes. Importantly, the
distortion of $I_c(H)$ is individual and each pattern in Fig.
\ref{fig:3} is clearly distinctive. As seen from Figs. \ref{fig:3}
(a-c) the $I_c(H)$ patterns are sensitive to the shape, Figs.
\ref{fig:2} (a) and \ref{fig:3} (a), the position, Figs.
\ref{fig:3} (b), and the amplitude, Figs. \ref{fig:3} (c), of
local field. Thus, the $I_c(H)$ contains an encrypted information
about local field distribution $B^*(x)$ and it should be possible
to extract it by proper analysis. This supports the statement of
Ref. \cite{Golod_2019a} that spatial resolution of a scanning
probe sensor based on a single planar junction is potentially not
limited by its size. Such a device could obviate a trade-off
problem between field sensitivity and spatial resolution inherent
for scanning SQUID sensors \cite{Kirtley_2016,Zeldov_2016}, as
mentioned in the Introduction.

\begin{figure*}[t]
    \includegraphics[width=0.98\textwidth]{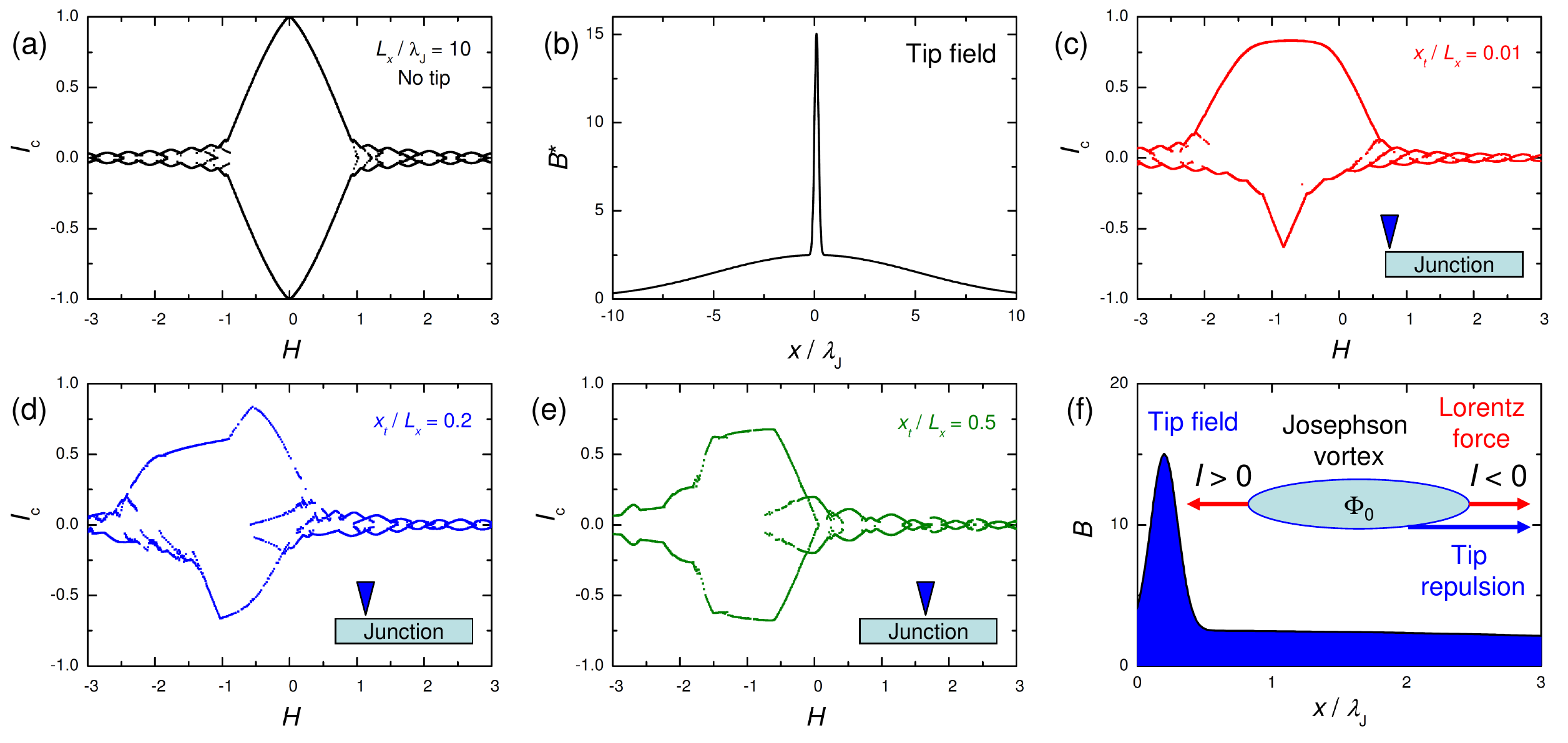}
     \caption{\label{fig:4} (Color online). Analysis of static characteristics of a long junction, $L_x=10\lambda_J$, subjected to a local
     nonuniform field from a tip of magnetic force microscope. (a) $I_c(H)$ modulation without a tip. (b) The assumed profile of the tip field.
     (c-e) $I_c(H)$ modulation patterns for different location of the tip, $x_t$, as indicated in the insets. (f) A sketch of interaction of a Josephson
     vortex with the tip placed at the left edge of the junction $x_t=0.01 L_x$. Tip induced local field and repulsion force are shown by the blue peak and arrow,
     respectively. Lorentz forces exerted by bias currents of different signs are depicted by red arrows. It is seen that the asymmetrically located local field
     creates a left-right asymmetry for JV motion and thus makes positive and negative critical currents dissimilar, as seen in panel (c). }
\end{figure*}

\subsection{III B. Long junctions}

For long JJ's, $L_x \gg \lambda_J$, screening of magnetic field by
the junction becomes significant. Simultaneously, Josephson
vortices (JV's) appear and start to affect junction properties. To
obtain $I_c$ with static $B^*$ either an ordinary differential
equation (14), or a dynamic partial differential equation (12)
with time-independent $\varphi^*$ should be solved with boundary
conditions, Eq. (13). Eq. (14) is solved by a finite difference
method with successive iterations and $I_c$ is determined as a
maximum bias current at which a solution converges. Eq. (12) is
integrated explicitly using a central difference approximation and
$I_c$ is determined using a threshold criterium for voltage. In
case of a significant nonlinearity of $\varphi^*(x)$, the
iterative solution of Eq. (14) may be quite sensitive to the
initial approximation. On the other hand, the damping term in Eq.
(12) allows less strict requirements to the initial approximation
and usually provides faster convergence because for the considered
here static case one can use a large $\alpha \simeq 1$ to speed up
calculations. Therefore, all simulations shown below are obtained
by solving partial differential Eq. (12).

Figure \ref{fig:4} (a) shows a simulated $I_c(H)$ pattern for a
long JJ with $L_x=10\lambda_J$, in the absence of local field
$B^*=0$. In contrast to the Fraunhofer $I_c(H)$ pattern for a
short JJ, Fig. \ref{fig:3} (a), it has a broad triangular central
lobe, corresponding to the Meissner state \cite{Barone}. Beyond it
JV's penetrate into the junction. Edge pinning of JV's, due to
interaction with their own images \cite{Golod_2019b}, leads to
metastability and multiply-valued $I_c$. Some of the metastable
states are seen in Figure \ref{fig:4} (a). To obtain those states
simulations are done by sweeping magnetic field back-and-forth in
different field intervals.

Next we consider the case with a local field. Here
we keep in mind another relevant case, when $B^*$ is induced by
the MFM tip \cite{Budakian_2019,Dremov_2019}. A standard MFM tip is covered
by a thin ferromagnetic layer. Therefore $B^*$ from the MFM sensor
has a sharp dipole-type peak, originating from the end of the tip,
and a broad background from the ferromagnetic layer at the
cantilever. To mimic it we approximate the tip-induced $B^*$ by
two Gaussian peaks: a narrow one with the width $\Delta
x_1=0.1\lambda_J$ containing a total flux of $0.5 \Phi_0$ and a
broad $\Delta x_2=5\lambda_J$ with the total flux $5 \Phi_0$, as
shown in Fig. \ref{fig:4} (b).

Fig. \ref{fig:4} (c-e) show simulated $I_c(H)$ patterns for the
same JJ with the MFM tip at different positions $x_t$ along
the junction, as indicated in insets. It is seen that the $I_c(H)$ in a long JJ is also distorted by the local field.
However, there are certain differences with respect to the short
junction case, Fig. \ref{fig:3}.

\subsection{III C. Asymmetry of $I_c(H)$ patterns}

From comparison of Figs. \ref{fig:3} and \ref{fig:4} (c-e) it can
be seen that local fields lead to qualitatively different
symmetries of $I_c(H)$ patterns for short and long JJ's. In the
absence of local field, $B^*=0$, the $I_c(H)$ patterns for both
short and long JJ's (with uniform parameters) are symmetric both
with respect to field and current directions. That is, positive,
$I_c^+$, and negative, $I_c^-$, critical currents are the same for
positive and negative fields: $I_c^+(H)= -I_c^-(H)$, $I_c^+(H) =
I_c^+(-H)$, see Figs. \ref{fig:3} (a) and \ref{fig:4} (a).

Local field $B^*\ne 0$ removes the space symmetry of the problem.
In all cases this removes the symmetry with respect to field
(space) inversion, $I_c(H)\ne I_c(-H)$. However, for short JJ's
the symmetry with respect to current (time) inversion is
preserved, $I_c^+(H) = -I_c^-(H)$. This occurs because for short
JJ's $I_c^+(H)$ and $I_c^-(H)$ correspond to maxima and minima of
the same integral of $\sin[(2\pi d_{eff} H/\Phi_0) x
+\varphi^*(x)+ \phi_0] dx$. Those are achieved at some constant
$\phi_0$ and $\phi_0+\pi$, respectively, thus leading to
$I_c^+(H)=-I_c^-(H)$.

In long JJ's local field removes the current reversal symmetry as
well, $I_c^+(H) \ne -I_c^-(H)$, see Figs. \ref{fig:4} (c) and (d).
This occurs because $I_c$ in long JJ's has a different nature: it
can be considered as a depinning current for JV's. Bias current
exerts a Lorentz force on JV's and leads to appearance of a
flux-flow state with finite voltage. For the considered geometry,
Fig. \ref{fig:1}, positive $J_b$ pushes JV's to the left and
negative - to the right, as indicated by red arrows in Figure
\ref{fig:4} (f). In the absence of local field, $B^*=0$, JV's are
pinned only at the edges of the junction due to attraction to
image antivortices \cite{Golod_2019b}. In this case $I_c^+$ and
$I_c^-$ correspond to depinning from the right and left edges,
respectively, which are equal in the absence of physical
nonuniformity of the JJ, as in Fig. \ref{fig:4} (a).

\begin{figure*}[t]
    \includegraphics[width=1.0\textwidth]{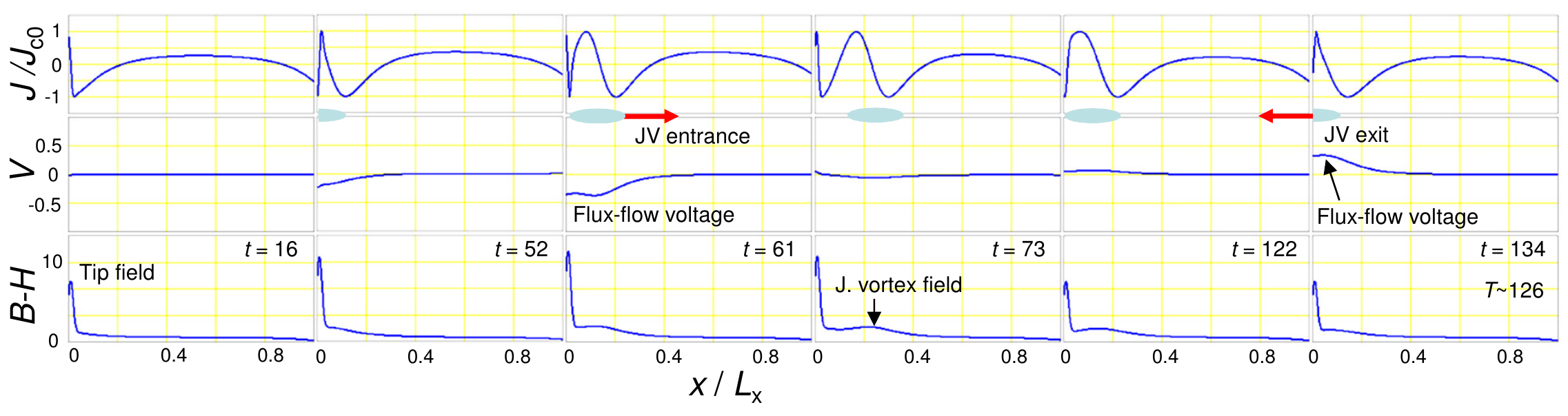}
     \caption{\label{fig:5} (Color online). Dynamics of a long junction $L_x=10\lambda_J$  at $H=-0.55$ induced by an oscillating MFM
     tip placed at the left edge of the junction $x_t=0.01 L_x$. Six time frames are shown (time increasing from left to right) within
     approximately one period of tip oscillation. For each frame the top panel shows a spatial distribution of the Josephson current, $\sin(\phi + \varphi^*)$, the middle
     panel - the voltage, $\partial (\phi + \varphi^*)/\partial t$, and the bottom panel - the inhomogeneous part of magnetic induction $B-H$. It is seen that the
     oscillating tip induces a shuttling motion of a single Josephson vortex, which enters and exits at the cite of the tip. }
\end{figure*}

The local inhomogeneous field with a finite gradient, $\partial
B^*/\partial x \ne 0$, exerts an additional magnetic force on the
JV, as indicated by the blue arrow in Fig. \ref{fig:4} (f), which
will remove the symmetry between left and right edges and lead to
$I_c^+(H) \ne I_c^-(H)$, except for the case when the symmetric
local field $B^*$ is placed symmetrically in the middle of the
junction, as in Fig. \ref{fig:4} (e). The local field creates also
an additional pinning cite for JV's inside the JJ. This leads to a
more profound metastability of $I_c(H)$ patterns as can be seen
from comparison of Figs. \ref{fig:4} (a) and (c-e).

So far we considered perfectly uniform JJ's. However, real JJ's
often contain some nonuniformities, e.g., they may have spatial
variation of intrinsic junction parameters, such as the critical
current density, electrode thickness, bias current density,
self-field effect, e.t.c. In this case the $I_c(H)$ pattern in the
absence of local field must only be centrosymmetric, $I_c^+(H)=
-I_c^-(-H)$ \cite{Krasnov_1997}. The latter is the consequence of
space-time symmetry: simultaneous reversal of field (space) and
current (time) is equivalent to flipping the junction upside-down,
which should not affect the output of the experiment. For
nonuniform junctions introduction of the local field $B^*$ removes
all sorts of symmetry $I_c^+(H) \ne -I_c^-(H)$ and $I_c^+(H) \ne
I_c^+(-H)$ even for short junctions (not shown).

\section{IV. Dynamic case}

The local field can be time-dependent, as for example in case of
MFM in the taping mode \cite{Dremov_2019}. The time-dependent
local field $B^*(t)$ provides an additional driving force for
junction dynamics, given by the last two terms in the right-hand
side of Eq. (\ref{SGnorm}). This can cause a flux-flow phenomenon
induced by the oscillating MFM tip, as recently reported
\cite{Dremov_2019}.

Figure \ref{fig:5} shows a time sequence of solutions of Eq.
(\ref{SGnorm}) with an oscillating MFM tip. Simulation parameters
correspond to Fig. \ref{fig:4} (c) : $L_x=10~\lambda_J$,
$x_t/L_x=0.01$ at $H=-0.55$. The damping parameter is
$\alpha=0.5$. The top panels show spatial distributions of the
Josephson current density, the middle panels - of voltage, and the
bottom panels - magnetization $B-H$. We assume that the tip is
oscillating harmonically, but the tip field is anharmonic due to
the non-linear distance dependence of the dipole-like tip field:
$B^*(t)=B^*(0)[1+a(1-\cos(\omega t)^3]$ with $a=0.5$ and
$\omega=0.05\omega_p$. Simulations are done for zero bias current
$j_b=0$. Thus, all the dynamics is induced solely by the
oscillating local field. Junction dynamics is periodic in time
with the period of tip oscillations $T=2\pi/\omega\simeq 125.7
\omega_p^{-1}$. Time sequences in Fig. \ref{fig:5} are shown for
approximately one period of tip oscillations.

The considered field $H=-0.55$ corresponds to a large $I_c^+$ and
a small $I_c^-$, see Fig. \ref{fig:4} (c). As discussed above, see
Fig. \ref{fig:4} (f), for a long junction the asymmetry $I_c^+(H)
\ne -I_c^-(H)$ makes left and right directions for motion of JV's
inequivalent. This in turn can lead to a variety of unusual
effects: the junction may act as a vortex-diode and rectify an
external periodic or aperiodic signal \cite{Krasnov_1997}. For the
case of a large $I_c^+$ and a small $I_c^-$ the easy direction of
JV motion is from left to right. From Fig. \ref{fig:5} it can be
seen that with increasing of the tip-induced field at $t\simeq
60\simeq T/2$ a Josephson vortex enters the junction from the left
side, where the tip is placed. After that it rapidly moves to the
right, inducing a significant negative flux-flow voltage, see the
middle panel at $t=61$. The JV penetrates to $x\sim 3 \lambda_J$,
see the frame at $t=73$. As the tip retracts the JV exits through
the left edge inducing a positive flux-flow voltage, see the frame
at $t=134$. A slight delay between tip and vortex oscillations is
caused by the viscosity of flux-flow motion due to a significant
damping $\alpha=0.5$.

Fig. \ref{fig:5} illustrates that the oscillating local field can
induce a flux-flow phenomenon in the junction. Depending on
parameters it can be the shuttling in/out vortex motion, as in
Fig. \ref{fig:5}, or a more complex ratchet-like unidirectional
motion with entrance of the JV from one side and exit from the
other side of the junction. An example of such ratchet-like motion
can be found in the Supplementary material to Ref.
\cite{Dremov_2019}. In that case every cycle four JV's enter a
junction from the left edge but only three leave from that side
while one exits through the right edge, thus creating a ratchet
effect, i.e. a net rectified unidirectional flux-flow motion
induced by a periodic (or aperiodic) perturbation
\cite{Krasnov_1997,Costabile_2001}. The back action of the
tip-induced flux-flow motion leads to an additional damping of MFM
tip oscillations, which can be detected in experiment. As
discussed in Ref. \cite{Dremov_2019}, this provides a mechanism
for detection of Josephson vortices by the MFM technique.

%\section{ Comparison with experiment}
\section{Conclusions}

To conclude, we derived and analyzed numerically equations
describing behavior of a Josephson junction in local inhomogeneous
magnetic field. As discussed in the Introduction, such situation
may have many different reasons and experimental realizations
\cite{VanHarlingen_1995,Tsuei_2000,Hilgenkamp_2003,Koshelev_2003,Goldobin_2008,Goldobin_2012,Bakurskiy_2016,Weides_2008,Birge_2012,Iovan_2014,Golovchanskiy_2016,Iovan_2017,Aarts_2017,Gaber_2005,Milosevic_2009,Finnemore_1994,Golod_2010,Golod_2015,Golod_2019b,Dremov_2019}.
It was demonstrated that time-dependent local field provides an
additional driving force, which may induce flux-flow type dynamics
in long junctions. This provides a mechanism for detection and
manipulation of Josephson vortices by taping-mode magnetic force
microscope \cite{Dremov_2019}. Local inhomogeneous field removes
the space-time symmetry of the junction and leads to a distortion
of $I_c(H)$ modulation patterns. Importantly, the distortion
uniquely depends on the spatial distribution of local field
$B^*(x)$ within the junction. Therefore, the information about
local field profile is encoded into the shape of the $I_c(H)$
pattern and may in principle be reconstructed using an appropriate
mathematical analysis. This strengthens an earlier argument that a
single planar junction can be advantageously used as a scanning
probe sensor \cite{Golod_2019a}. The field sensitivity of such
sensor would depend on the area, similar to SQUID, but the spatial
resolution would not be limited by the junction size. Therefore, a
planar junction sensor can obviate the trade-off problem between
field sensitivity and spatial resolution inherent in scanning
SQUID microscopy.

\begin{acknowledgments}
Acknowledgments:  I am grateful to T. Golod for assistance with
assembling Fig. \ref{fig:3} and to V. V. Dremov, R. A.
Hovhannisyan, S. Yu. Grebenchuk and V. S. Stolyarov for
stimulating discussions about MFM experiments. The work was
supported by the Russian Science Foundation, Grant No.
19-19-00594.
%and the European Union H2020-WIDESPREAD-05-2017-Twinning project SPINTECH under Grant Agreement No.810144.
The manuscript was accomplished
during a sabbatical semester at MIPT, supported by the
5-top-100 program.
\end{acknowledgments}.

\end {document}